\documentclass[sigconf]{acmart}

\usepackage{graphicx}
	\graphicspath{{graphics/}}
	\DeclareGraphicsExtensions{.pdf,.jpeg,.png}
\usepackage[percent]{overpic}
\usepackage{algpseudocode}
\usepackage[caption=false,font=normalsize,labelfont=sf,textfont=sf]{subfig}
\usepackage{enumitem}
\usepackage{tikz}

\copyrightyear{2017} 
\acmYear{2017} 
\setcopyright{acmlicensed}
\acmConference{DEBS '17}{June 19-23, 2017}{Barcelona, Spain}\acmPrice{15.00}\acmDOI{10.1145/3093742.3095103}
\acmISBN{978-1-4503-5065-5/17/06}

\begin{document}
\title{Grand Challenge: StreamLearner -- Distributed Incremental \\ Machine Learning on Event Streams}

\author{Christian Mayer, Ruben Mayer, and Majd Abdo}
\affiliation{%
  \institution{{Institute for Parallel and Distributed Systems }}
  \institution{{University of Stuttgart, Germany}}
}
\email{firstname.lastname@ipvs.uni-stuttgart.de}

\renewcommand{\shortauthors}{Mayer et al.}

\begin{abstract}
Today, massive amounts of streaming data from smart devices need to be analyzed automatically to realize the Internet of Things. The Complex Event Processing (CEP) paradigm promises low-latency pattern detection on event streams.
However, CEP systems need to be extended with Machine Learning (ML) capabilities such as online training and inference in order to be able to detect fuzzy patterns (e.g. outliers) and to improve pattern recognition accuracy during runtime using incremental model training.
In this paper, we propose a distributed CEP system denoted as StreamLearner for ML-enabled complex event detection. The proposed programming model and data-parallel system architecture enable a wide range of real-world applications and allow for dynamically scaling up and out system resources for low-latency, high-throughput event processing.
We show that the DEBS Grand Challenge 2017 case study (i.e., anomaly detection in smart factories) integrates seamlessly into the StreamLearner API. Our experiments verify scalability and high event throughput of StreamLearner.

\end{abstract}




\maketitle

\begin{tikzpicture}
\begin{scope}[overlay]
\node[text width=20cm] at ([yshift=-5.0cm]current page.south) {(c) Owner 2017. This is the authors' version of the work. It is posted here for your personal use. Not for redistribution. \newline The definitive version is published in Proceedings of ACM International Conference on Distributed and Event-Based \newline Systems 2017 (DEBS '17), http://dx.doi.org/10.1145/3093742.3095103.};
\end{scope}
\end{tikzpicture}

\section{Introduction and Background}
\label{sec:introduction}

In recent years, the surge of \emph{Big Streaming Data} being available from sensors \cite{krishnan2014activity}, social networks \cite{Mayer2016GraphCEP}, and smart cities \cite{batty2013big}, has led to a shift of paradigms in data analytics throughout all disciplines. Instead of \emph{batch-oriented} processing \cite{Dean:2008:MSD:1327452.1327492,Malewicz2010Pregel,Mayer2016GrapH}, \emph{stream-oriented} data analytics \cite{cugola2012processing} is becoming the gold standard. This has led to the development of scalable \emph{stream processing systems} that implement the relational query model of relational data base management systems (RDBMS) as \emph{continuous queries} on event streams \cite{Arasu:2006:CCQ:1146461.1146463}, and \emph{Complex Event Processing systems} that implement pattern matching on event streams \cite{Cugola:2010:TFD:1827418.1827427}.

Query-driven stream processing, however, demands a domain expert to specify the analytics logic in a deterministic query language with a query that exactly defines which input events are transformed into which output events by an \emph{operator}. However, an explicit specification is not always possible, as the domain expert might rather be interested in a more abstract query such as ``\textit{Report me all anomalies that molding machine 42 experiences on the shopfloor.}'' In this example, it is infeasible to explicitly specify all event patterns that can be seen as an \textit{anomaly}.

There have been different proposals how to deal with this issue. EP-SPARQL employs background ontologies to empower (complex) event processing systems with stream reasoning \cite{Anicic:2011:EUL:1963405.1963495} -- while focusing on the SPARQL query language.
On the other hand, several general-purpose systems for stream processing exist such as Apache Kafka \cite{kreps2011kafka}, Apache Flink \cite{carbone2015apache}, Apache Storm \cite{toshniwal2014storm}, Apache Spark Streaming \cite{zaharia2013discretized}.
Although these systems are powerful and generic, they are not tailored towards parallel and scalable incremental model training and inference on event streams.

At the same time, an increasing body of research addresses incremental (or online) updates of Machine Learning (ML) models: there are incremental algorithms for all kinds of ML techniques such as support vector machines \cite{cauwenberghs2001incremental}, neural networks \cite{furao2007enhanced}, or Bayesian models \cite{wilson2010bayesian}.
Clearly, a stream processing framework supporting intuitive integration of these algorithms would be highly beneficial -- saving the costs of hiring expensive ML experts to migrate these algorithms to the stream processing systems.

In this paper, we ask the question: how can we combine event-based stream processing (e.g., for pattern recognition) with powerful Machine Learning functionality (e.g., to perform anomaly detection) in a way that is compatible with existing incremental ML algorithms?
We propose the distributed event processing system StreamLearner that decouples expertise of Machine Learning from Distributed CEP using a general-purpose modular API. In particular, we provide the following contributions.

\begin{itemize}
	\item An architectural design and programming interface for data-parallel CEP that allows for easy integration of existing incremental ML algorithms (cf. Section~\ref{sec:streamlearner}).
	\item An algorithmic solution to the problems of incremental K-Means clustering and Markov model training in the context of anomaly detection in smart factories (cf. Section~\ref{sec:casestudy}).
	\item An evaluation showing scalability of the StreamLearner architecture and throughput of up to 500 events per second using our algorithms for incremental ML model updates (cf. Section~\ref{sec:evaluation}).
\end{itemize}

%
%
%
%
%
%
%
%
%
%
\section{Challenges and Goals}
\label{sec:challenges}

Machine Learning algorithms train a model using a given set of training data, e.g., building clusters, and then apply the trained model to solve problems, e.g., classifying unknown events.
In the course of streaming data becoming available from sensors, models need to be dynamically adapted. That means, that new data is taken into account in the learned model, while old data ``fades out'' and leaves the model as it becomes irrelevant. This can be modeled by a \emph{sliding window} over the incoming event streams: Events within the window are relevant for the model training, whereas events that fall out of the window become irrelevant and should not be reflected in the model any longer.
Machine Learning on sliding windows is also known as \textit{non-stationary Machine Learning}, i.e., the problem of keeping a model updated as the underlying streaming data generation ``process'' underlies a changing probability distribution.
To adapt the ML model \textit{online}, there are different possibilities. For instance, incremental algorithms change the model in a step-by-step fashion. The challenge in doing so is to support incremental processing -- i.e., streaming learning. The model should not be re-built from scratch for every new window, but rather incrementally be updated with new data while old data is removed.

Another challenge in ML in streaming data is that data from different streams might lead to independent models. For instance, data captured in one production machine might not be suitable to train the model of another production machine. The challenge is to determine which independent models shall be built based on which data from which incoming event streams. Further, the question is how to route the corresponding events to the appropriate model. When these questions are solved, the identified machine learning models can be built in parallel -- enabling scalable, low-latency, and high-throughput stream processing.

\section{StreamLearner}
\label{sec:streamlearner}

In this section, we first give an overview about the StreamLearner architecture, followed by a description of the easy-to-use API for incremental machine learning and situation inference models.



\begin{figure}
	\includegraphics[width=\linewidth]{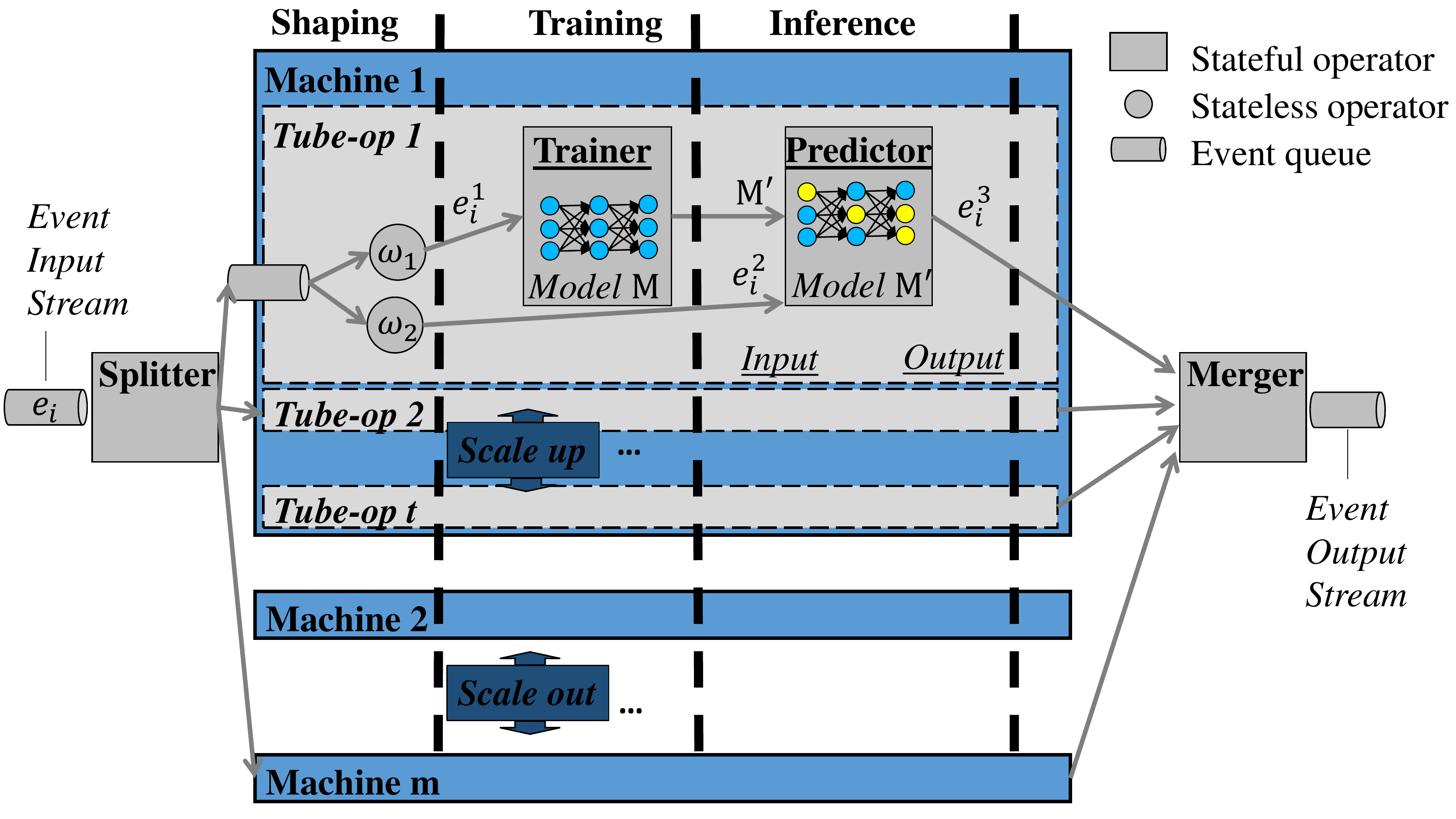}
	\caption{System Architecture}
	\label{fig:architecture}
\vspace{-0.4cm}
\end{figure}

\subsection{System Overview}


The architecture of StreamLearner is given in Figure~\ref{fig:architecture}. In order to parallelize ML-based computation, we have extended the split-process-merge architecture of traditional event-based systems \cite{Mayer2016GraphCEP, 7024105, Mayer2016Minimizing}. The splitter receives events via the event input stream and forwards them to independent processing units, denoted as \textit{tube-ops}, according to its splitting logic. Each tube-op atomically performs ML-based incremental stream processing by reading an event from the in-queue, processing the event, and forwarding the output event to the merger. The merger decides about the final events on the event output stream (e.g. sorts the events from the different tube-ops by timestamp to provide a consistent ordering of the event output stream). Due to the independent processing of events, the architecture supports both, scale-up operations by spawning more threads per machine and scale-out operations by adding more machines.

Each tube-op processes an event in three phases: shaping, training, and inference. In the shaping phase, it performs stateless preprocessing operations $\omega_1$ and $\omega_2$ (denoted as shaper) to transform the input event into appropriate formats. In the training phase, the stateful trainer module incrementally updates the model parameters of model $M$ (e.g. a neural network in Figure~\ref{fig:architecture}) according to the user-specified model update function. In the inference phase, the updated model and the preprocessed event serve as an input for the stateful predictor performing a user-defined inference operation and transforming the updated model and the input event to an output event with the model-driven prediction.

Note that the StreamLearner API does not restrict application programmers to perform training and inference on \textit{different event data}. Hence, application programmers are free to use either \textit{disjoint subsets}, or \textit{intersecting subsets}  of events in the stream for training and inference.
Although it is common practice in ML to separate data that is used for training and inference, we still provide this flexibility, as in real-world streams we might use some events for both, incorporating changing patterns into the ML model \textit{and} initiating an inference event using the predictor.
However, the application programmer can also separate training and inference data by defining the operators in the tube-op accordingly (e.g. generating a \textit{dummy event} as input for the predictor to indicate that no inference step should be performed). Furthermore, the application programmer can also specify whether the training should happen before inference or vice versa.

\subsection{Programming Model}

The application programmer specifies the following functions in order to use the StreamLearner framework in a distributed environment.

\subsubsection{Splitter}
Given an event $e_i$, the application programmer defines a stateful splitting function $split(e_i)$ that returns a tuple $(mid, tid, e_i)$ defining the tube-op $tid$ on machine $mid$ that receives event $e_i$.

\subsubsection{Shaping}
The stateless shaper operations $\omega_1(e_i)$ and $\omega_2(e_i)$ return modified events $e_i^1$ and $e_i^2$ that serve as input for the trainer and the predictor module. The default shaper performs the identity operation.

\subsubsection{Trainer}
The stateful trainer operation $trainer(e_i^1)$ returns a reference to the updated model object $M'$. The application programmer can use any type of machine learning model as long as the model can be used for inference by the predictor. If the model $M$ remains unchanged after processing event $e_i^1$, the trainer must return a reference to the unchanged model $M$ in order to trigger the predictor for each event.
StreamLearner performs a delaying strategy when the application programmer prefers inference \textit{before} learning. In this case, the tube-op first executes the predictor on the old model $M$ and executes the trainer afterwards to update the model.

\subsubsection{Predictor}
The stateful predictor receives a reference to model $M'$ and input (event) $e_i^2$ and returns the predicted event $e_i^3=predictor(M',e_i^2)$.

\subsubsection{Merger}
The stateful merger receives predicted output events from the tube-ops and returns a sequence of events that is put to the event output stream, i.e., $merger(e_i^3)=f(e_0^3,...,e_j^3,...,e_i^3)$ for $j<i$ and any function $f$. Any aggregator function, event ordering scheme, or filtering method can be implemented by the merger.

\section{Case Study: Anomaly Detection in Smart Factories}
\label{sec:casestudy}

In this section, we exemplify usage of our StreamLearner API based on a realistic use case for data analytics posed by the DEBS Grand Challenge 2017\footnote{http://www.debs2017.org/call-for-grand-challenge-solutions/} \cite{DEBSGC2017}.

\subsection{Problem Description}
In smart factories, detecting malfunctioning of production machines is crucial to enable automatic failure correction and timely reactions to bottlenecks in the production line.
The goal of this case study is to detect anomalies, i.e., abnormal sequences of sensor events quantifying the state of the production machines. In particular, the input event stream consists of events transporting measurements from a set of production machines $P$ to an anomaly detection operator.
The events are created by the set of sensors $S$ that monitor the production machines.
We include the time stamps of each measured sensor event by defining a set of discrete time steps $DT$. Each event $e_i=(p_i, d_i, s_i, t_i)$ consists of a production machine id $p_i \in P$ that was monitored, a numerical data value $d_i \in \mathbb{R}$ quantifying the state of the production machine (e.g. temperature, pressure, failure rate), a sensor with id $s_i \in S$ that has generated the event, and a time stamp $t_i \in DT$ storing the event creation time.

\begin{figure}
	\includegraphics[width=\linewidth, clip=true, trim=0pt 140pt 0pt 0pt]{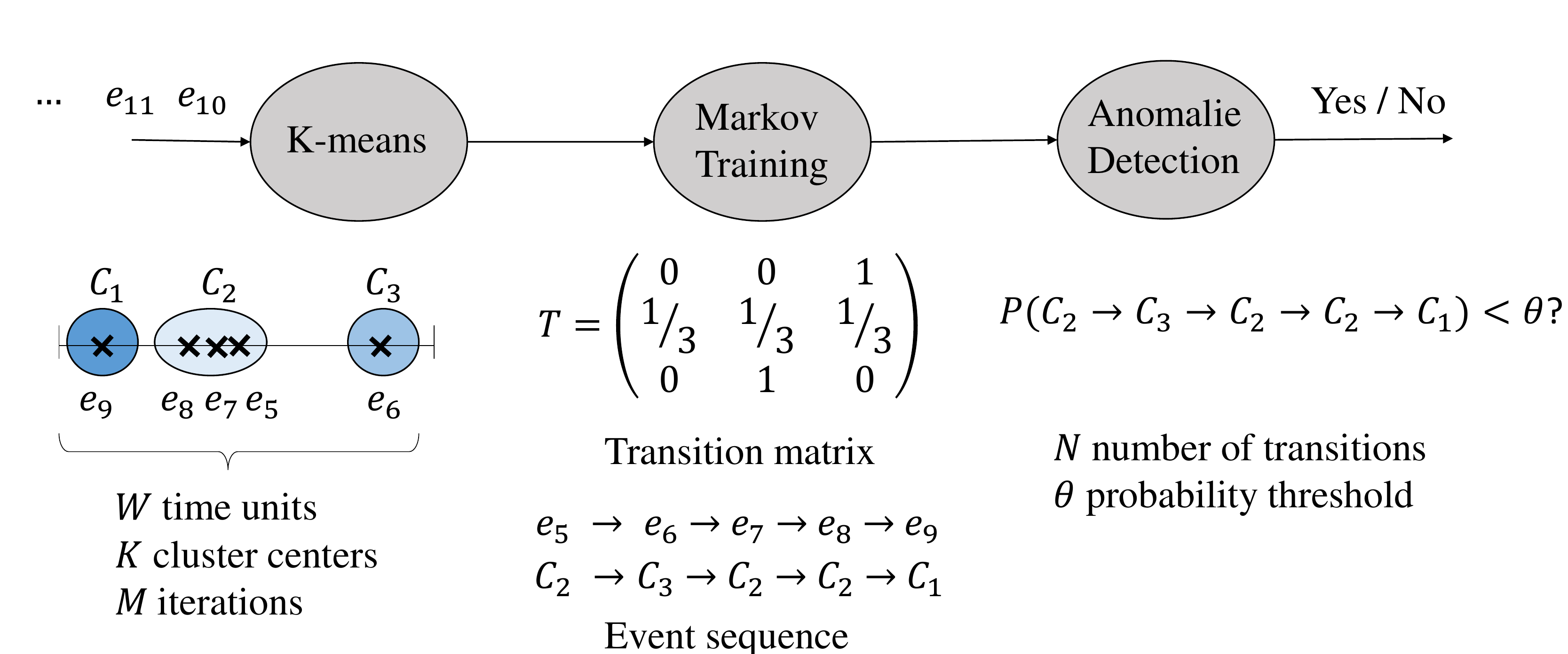}
	\caption{Case study \textit{anomaly detection} in smart factories.}
	\label{fig:caseStudyProblem}
\vspace{-0.5cm}
\end{figure}

The anomaly detection operator has to pass three stages for each event-generating sensor (cf. Figure~\ref{fig:caseStudyProblem}).

First, it collects all events $e_i$ that were generated within the last $W$ time units (denoted as \textit{event window}) and clusters the events $e_i$ using the $K$-means algorithm on the numerical data values $d_i$ for at least $M$ iterations. The standard $K$-means algorithm iteratively assigns each event in the window to its closest cluster center (with respect to euclidean distance) and recalculates each cluster center as the centroid of all assigned events' numerical data values (in the following we do not differentiate between events and their data values).
In the figure, there are five events $e_5, e_6, e_7, e_8, e_9$ in the event window that are clustered into three clusters $C_1,C_2,C_3$.
With this method, we can characterize each event according to its \textit{state}, i.e., the cluster it is assigned to.

Second, the operator trains a first-order Markov model in order to differentiate normal from abnormal event sequences. A Markov model is a state diagram, where a probability value is associated to each state transition. The probability of a state transition depends only on the current state and \textit{not} on previous state transitions (independence assumption). 
These probabilities are maintained in a transition matrix $T$ using the following method:
(i) The Markov model consists of $K$ states, one state for each cluster. Each event is assumed to be in the state of the cluster it is assigned to.
(ii) The events are ordered with respect to their time stamp -- from oldest to youngest. Subsequent events are viewed as state transitions. In Figure~\ref{fig:caseStudyProblem}, the events can be sorted as $[e_5, e_6, e_7, e_8, e_9]$. The respective state transitions are $C_2 \to C_3 \to C_2 \to C_2 \to C_1$.
(iii) The transition matrix contains the probabilities of state transitions between any two states, i.e., cluster centers. The probability of two subsequent events being in cluster $C_i$ and transition into cluster $C_j$ for all $i,j \in \{1,...,K\}$ is the relative number of these observations. For example the probability of transition from state $C_2$ to state $C_1$ is the number of events in state $C_2$ that transition to state $C_1$ divided by the total number of transitions from state $C_2$, i.e., $P(C_1|C_2)=\frac{\#C_2 \to C_1}{\#C_2 \to \star}=1/3$.

Third, an anomaly is defined using the \textit{probability of a sequence of observed transitions with length $N$}. In particular, if a series of unlikely state transitions is observed, i.e., the total sequence probability is below the threshold $\Theta$, an event is generated that indicates whether an anomaly has been found.
The probability of the sequence can be calculated by breaking the sequence into single state transitions, i.e., in Figure~\ref{fig:caseStudyProblem}, $P(C_2 \to C_3 \to C_2 \to C_2 \to C_1)=P(C_2 \to C_3) P(C_3 \to C_2) P(C_2 \to C_2) P(C_2 \to C_1)$. Using the independence assumption of Markov models, we can assign a probability value to each sequence of state transition and hence quantify the likelihood.



\subsection{Formulating the Problem in the StreamLearner API}
The scenario fits nicely into the StreamLearner API: for each sensor, an independent ML model is subject to incremental training and inference steps. Therefore, each thread in the StreamLearner API is responsible for all observations of a single sensor enabling StreamLearner to monitor multiple sensors in parallel.

\subsubsection{Splitter}
The splitter receives an event $e_i=(p_i, d_i, s, t_i)$ and assigns the event exclusively to the thread that is responsible for sensor $s$ (or initiates creation of this responsible thread if it does not exist yet). It uses a simple hash map assigning sensor ids to thread ids to provide thread resolution with constant time complexity during processing of the input event stream. With this method, we break the input stream into multiple independent sensor event streams (one stream per sensor). 

\subsubsection{Shapers}
Shapers $\omega_1$ and $\omega_2$ are simply identity operators that pass the event without changes to the respective training or prediction modules.

\subsubsection{Trainer}
The trainer maintains and updates the model in an incremental fashion.
The model is defined via the transition matrix $T$ that is calculated using K-means clustering and the respective state transition sequence.

\paragraph{Incremental K-Means} The goal is to iteratively assign each event to the closest cluster center and recalculate the cluster center as the centroid of all assigned events. The standard approach is to perform $M$ iterations of the K-means clustering algorithm for all events in the event window when triggered by the arrival of a new event. However, this method results in suboptimal runtime due to unnecessary computations that arise in practical settings:

\begin{itemize}
	\item A single new event in the event window will rarely have a global impact to the clustering. In particular, most assignments of events to clusters remain unchanged after adding a new event to the event window. Therefore, the brute-force method of full reclustering can result in huge computational redundancies.
	\item Performing $M$ iterations is unnecessary, if the clustering has already converged in an earlier iteration $M'<M$. Clearly, we should terminate the algorithm as fast as possible.
	\item The one-dimensional K-means problem is fundamentally easier than the standard NP-hard K-means problem: an optimal solution can be calculated in polynomial time $\mathcal{O}(n^2K)$ for fixed number of clusters $K$ and number of events in the window $n$ \cite{wang2011ckmeans, mahajan2009planar}. Therefore, using a general-purpose K-means algorithm that supports arbitrary dimensionality can result in unnecessary overhead (the trade-off between generality, performance, and optimality).
\end{itemize}

\begin{figure}
	\includegraphics[width=\linewidth, clip=true, trim=0pt 250pt 0pt 0pt]{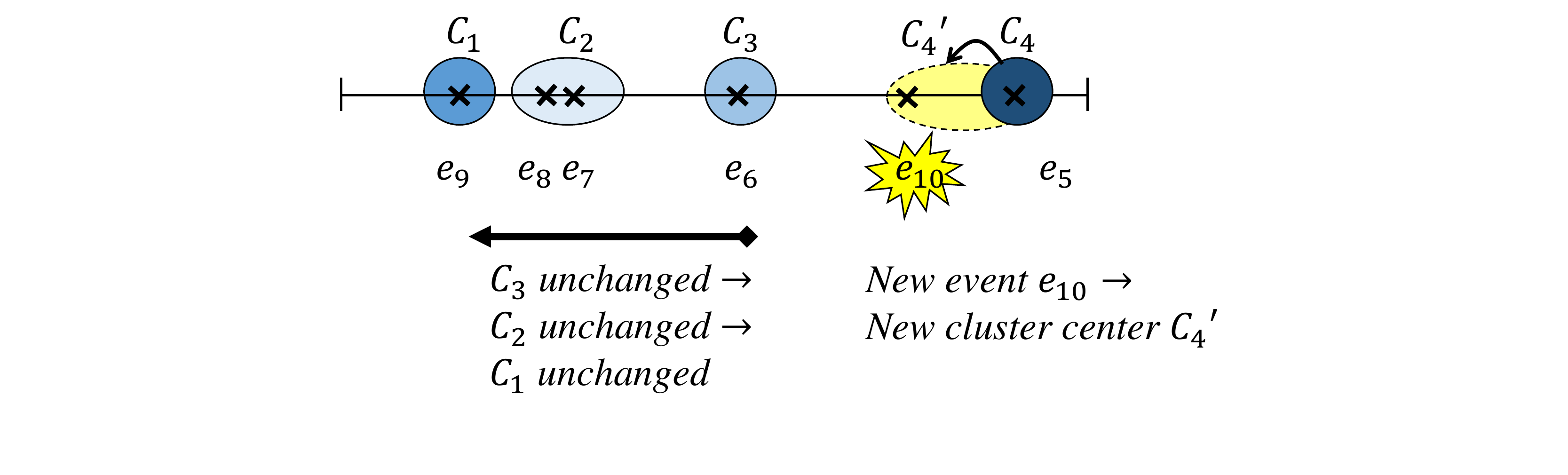}
	\caption{Saving computation time in K-Means.}
	\label{fig:kmeans}
\end{figure}

This is illustrated in Figure~\ref{fig:kmeans}. There are four clusters $C_1,...,C_4$ and events $e_5,...,e_9$ in the event window. A new event $e_{10}$ is arriving. Instead of recomputation of the whole clustering in each iteration, i.e., calculating the distance between each event and cluster center, we \textit{touch} only events that are potentially affected by a change of the cluster centers. For example, event $e_{10}$ is assigned to cluster $C_4$ which leads to a new cluster center $C_4'$. However, the next closest event $e_6$ (left side) keeps the same cluster center $C_3$. Our basic reasoning is that each event on the left side of the unchanged event $e_6$ keeps \textit{its} cluster center as there can be no disturbance in the form of changed cluster centers left-hand of $e_6$ (only a cascading cluster center shift is possible as $C_4 \geq C_3 \geq C_2 \geq C_1$ in any phase of the algorithm). A similar argumentation can be made for the right side and also for the removal of events from the window.

This idea heavily utilizes the possibility of sorting cluster centers and events in the one-dimensional space. It reduces average runtime of a single iteration of K-means as in many cases only a small subset of events has to be accessed. Combined with the optimization of skipping further computation after convergence in iteration $M'<M$, incremental updates of the clustering can be much more efficient than naive reclustering. The incremental one-dimensional clustering method is in the same complexity class as naive reclustering as in the worst case, we have to reassign all events to new clusters (the sorting of events takes only logarithmic runtime complexity in the event window size per insertion of a new event -- hence the complexity is dominated by the K-means computation). 

\paragraph{Markov Model}
The Markov model is defined by the state transition matrix $T$. Cell $(i,j)$ in the transition matrix $T$ is the probability of two subsequent events to transition from cluster $C_i$ (the first event) to cluster $C_j$ (the second event). Semantically, we \textit{count} the number of state transitions in the event window to determine the relative frequency such that the row values in $T$ sum to one. Instead of complete recomputation of the whole matrix, we only recalculate the rows and columns of clusters that were subject to any change in the K-means incremental clustering method. This ensures that all state transitions are reflected in the model while saving computational overhead. A reference to the new model $T$ is handed to the predictor method that performs inference on the updated model as presented in the following.

\subsubsection{Predictor}
The predictor module applies the inference step on the changed model for each incoming event.
In this scenario, inference is done via the Markov model (i.e., the transition matrix $T$) to determine whether an anomaly was detected or not.
We use the transition matrix to assign a probability value to a sequence of events with associated states (i.e., cluster centers).
The brute-force method would calculate the product of state transition probabilities for each sequence of length $N$ and compare it with the probability threshold $\Theta$. However, this leads to many redundant computations for subsequent events.

\begin{figure}
	\includegraphics[width=\linewidth, clip=true, trim=0pt 0pt 0pt 0pt]{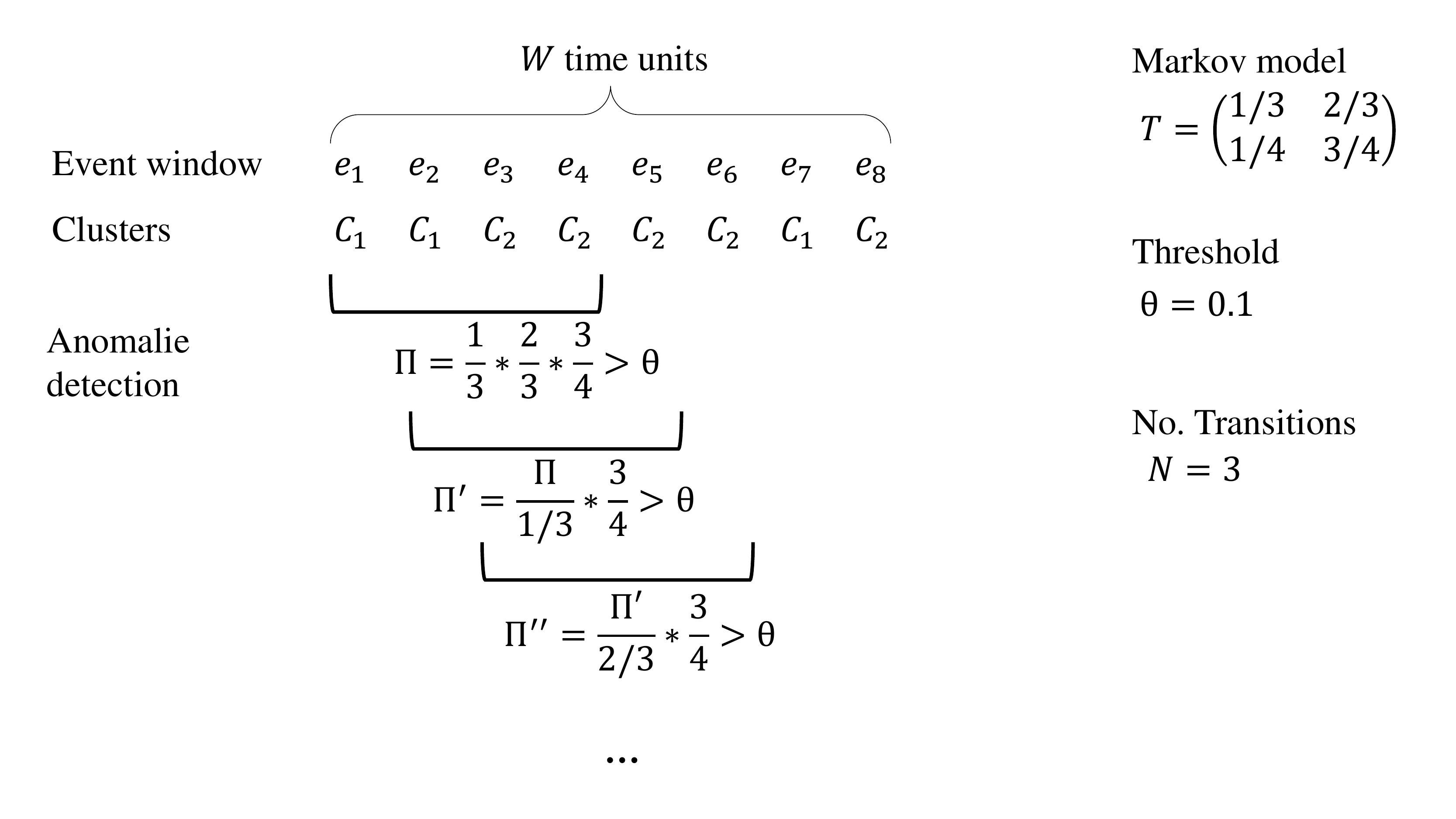}
	\caption{anomaly detection.}
	\label{fig:anomalyDetection}
\end{figure}

We present an improved incremental method in Figure~\ref{fig:anomalyDetection}. The event window consists of events $e_1,...,e_8$ sorted by time stamps. Each event is assigned to a cluster $C_1$ or $C_2$ resulting in a series of state transitions. We use the transition matrix of the Markov model to determine the probability of each state transition.
We calculate the probability of the state transition sequence as the product of all state transitions (the state independence property of Markov models). For instance the probability $\Pi$ of the first three state transitions is $\Pi=P(C_1|C_1)*P(C_2|C_1)*P(C_2|C_2)=1/3*2/3*3/4=1/4$ which is larger than the threshold $\Theta=0.1$. Now we can easily calculate the probability of the next state transition sequence of length $N$ by dividing by the first transition probability of the sequence (i.e., $P(C_1|C_1)=1/3$) and multiplying with the probability of the new state transition (i.e., $P(C_2|C_2)=3/4$). Hence, the total probability $\Pi'$ of the next state transition sequence is $\Pi'=\frac{\Pi}{1/3}*3/4=9/16>\Theta$. This method reduces the number of multiplications to $N+2(W-N)$ rather than $N(W-N)$. Finally, the predictor issues an anomaly detection event to the merger (Yes/No).

\subsubsection{Merger}
The merger sorts all anomalies events w.r.t. time stamp to ensure a consistent output event stream using the same procedure as in GraphCEP \cite{Mayer2016GraphCEP}. This method ensures a monotonic increase of event time stamps in the output event stream.




\begin{figure*}
	\centering
	%
	\subfloat[Absolute Throughput.]{\label{fig:scalabilityNotebook}	 \includegraphics[width=0.33\linewidth, clip=true, trim=0pt 0pt 0pt 0pt]{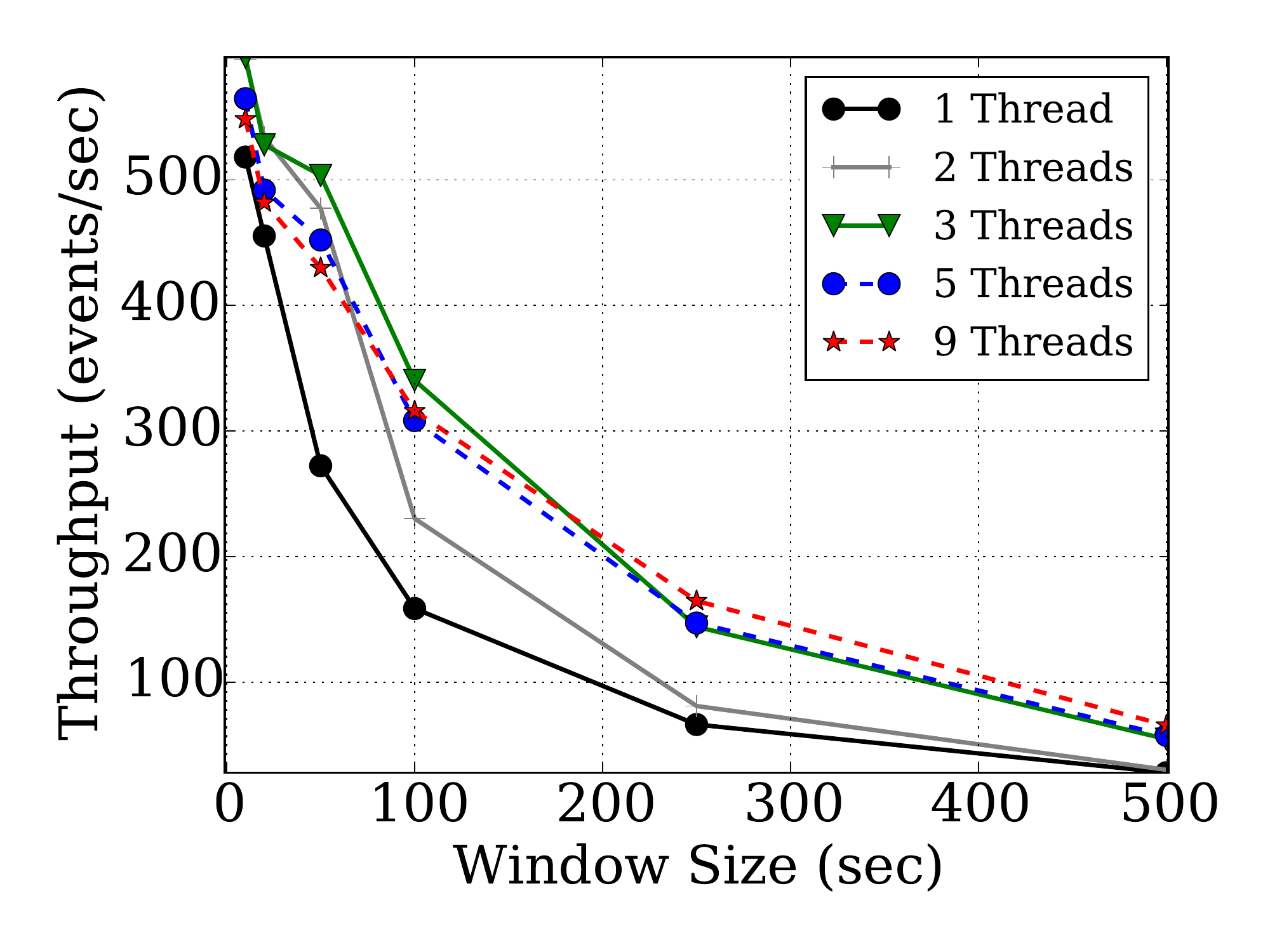}}
	%
	\subfloat[Normalized Throughput.]{\label{fig:scalabilityNotebookNormalized}	 \includegraphics[width=0.33\linewidth, clip=true, trim=0pt 0pt 0pt 0pt]{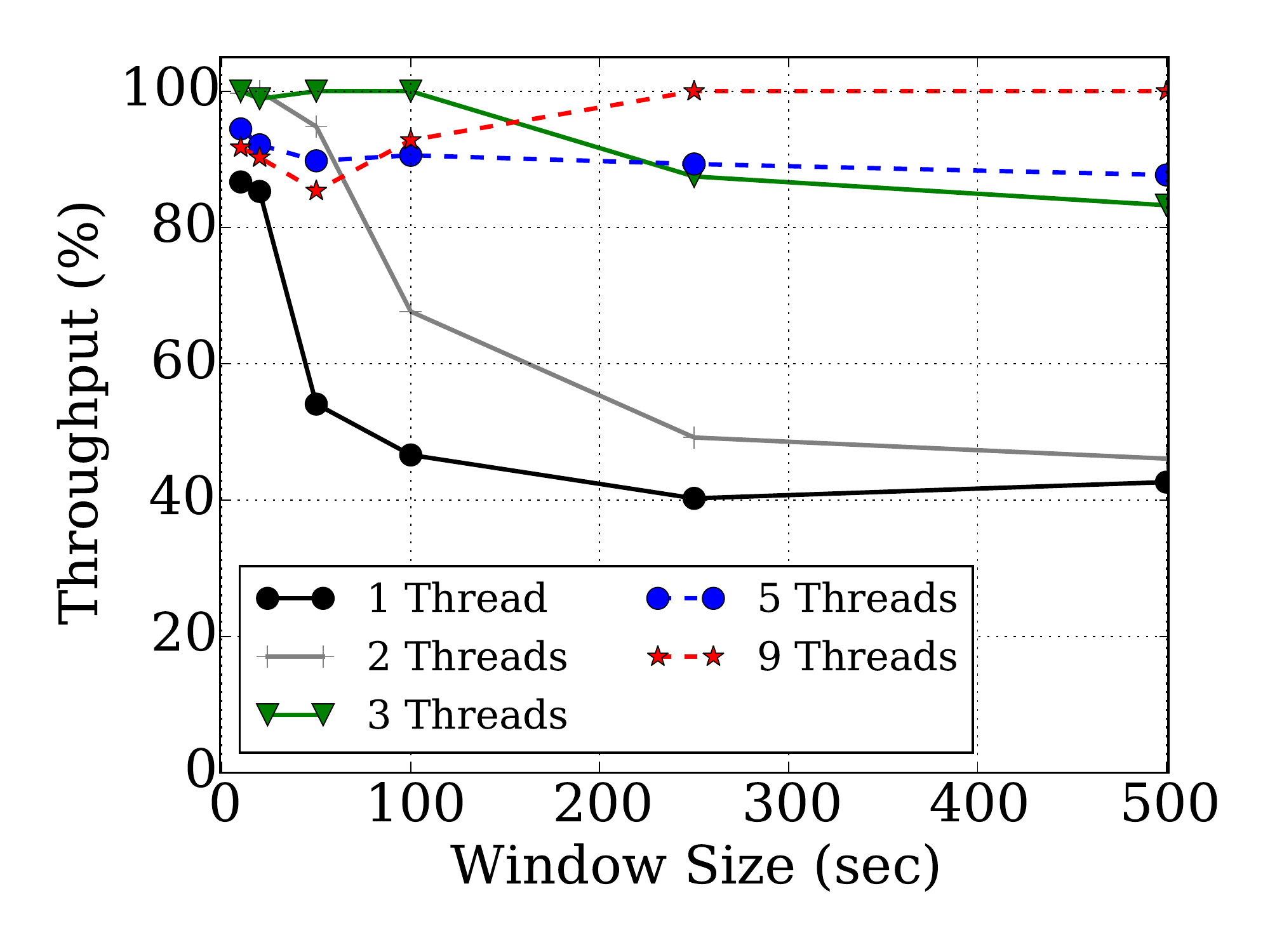}}
	%
	\subfloat[Scalability.]{\label{fig:scalabilityNotebookBar}	 \includegraphics[width=0.33\linewidth, clip=true, trim=0pt 0pt 0pt 0pt]{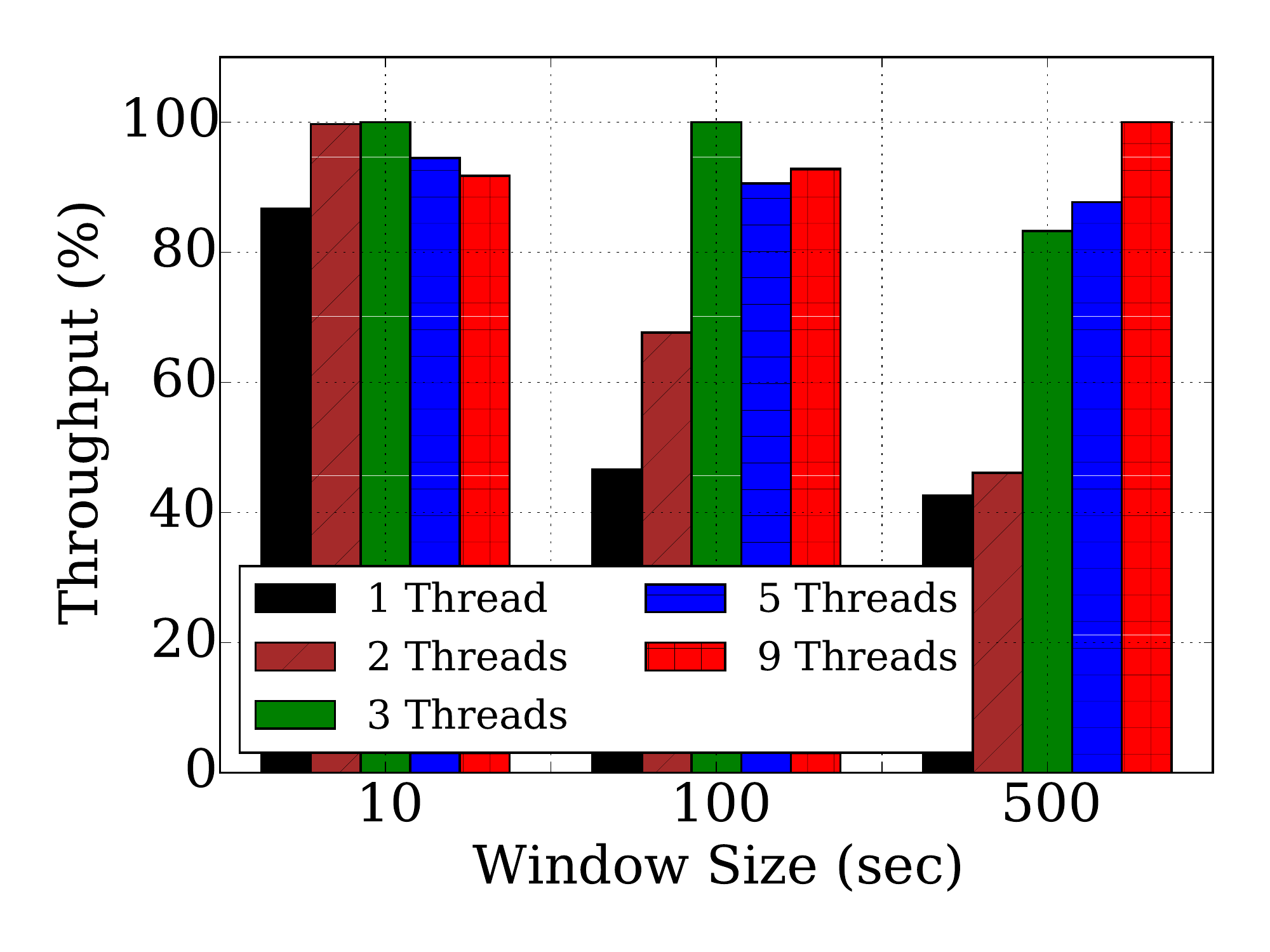}}
\vspace{-8pt}
	\caption{Throughput evaluations for different window sizes $W$ on notebook.}
	\label{fig:evalsNotebook}
	\vspace{-20pt}
\end{figure*}

\begin{figure*}
	\centering
	%
	\subfloat[Absolute Throughput.]{\label{fig:scalabilityOctopus}	 \includegraphics[width=0.33\linewidth, clip=true, trim=0pt 0pt 0pt 0pt]{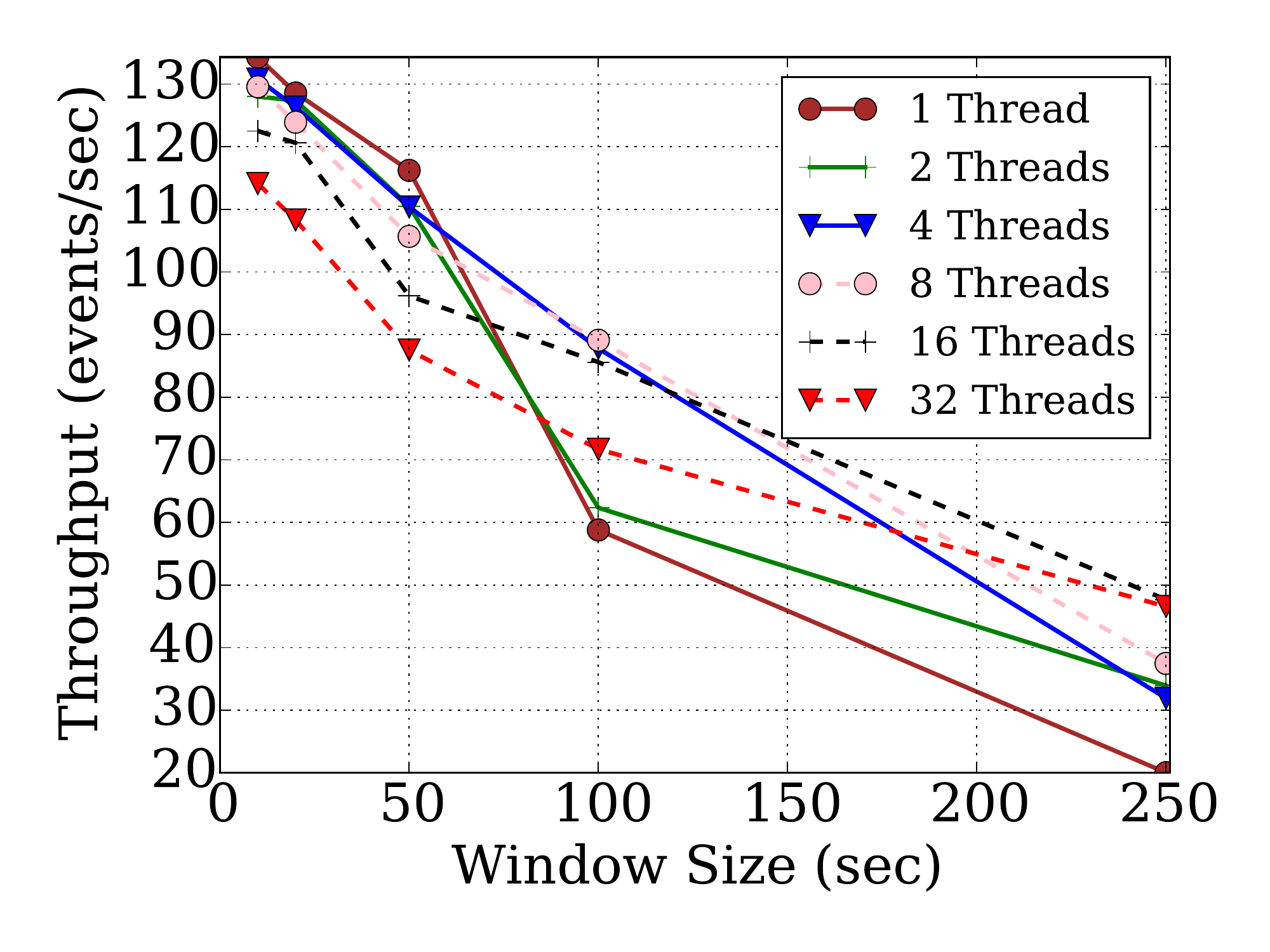}}
	%
	\subfloat[Normalized Throughput.]{\label{fig:scalabilityOctopusNormalized}	 \includegraphics[width=0.33\linewidth, clip=true, trim=0pt 0pt 0pt 0pt]{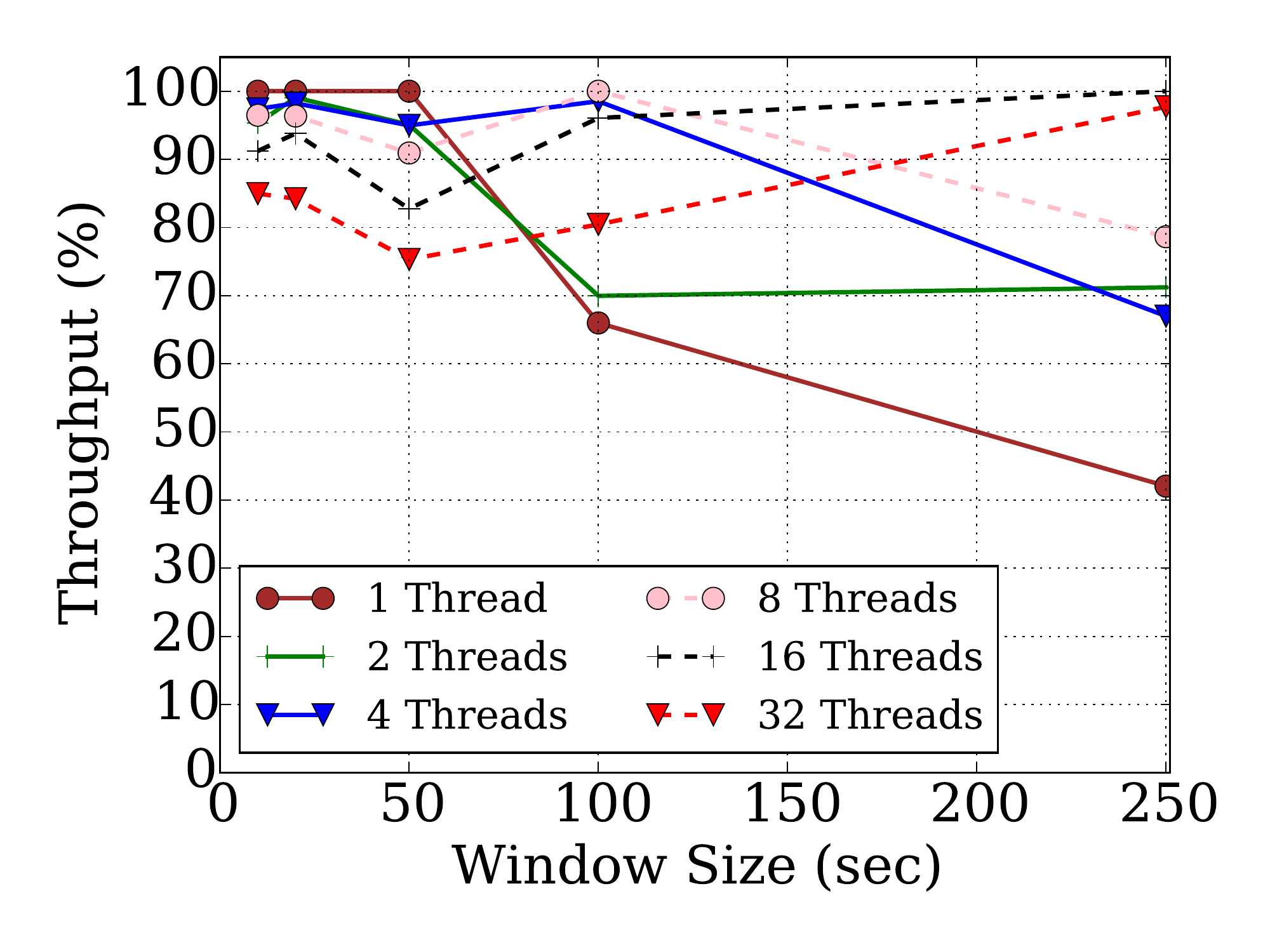}}
	%
	\subfloat[Scalability.]{\label{fig:scalabilityOctopusBar}	 \includegraphics[width=0.33\linewidth, clip=true, trim=0pt 0pt 0pt 0pt]{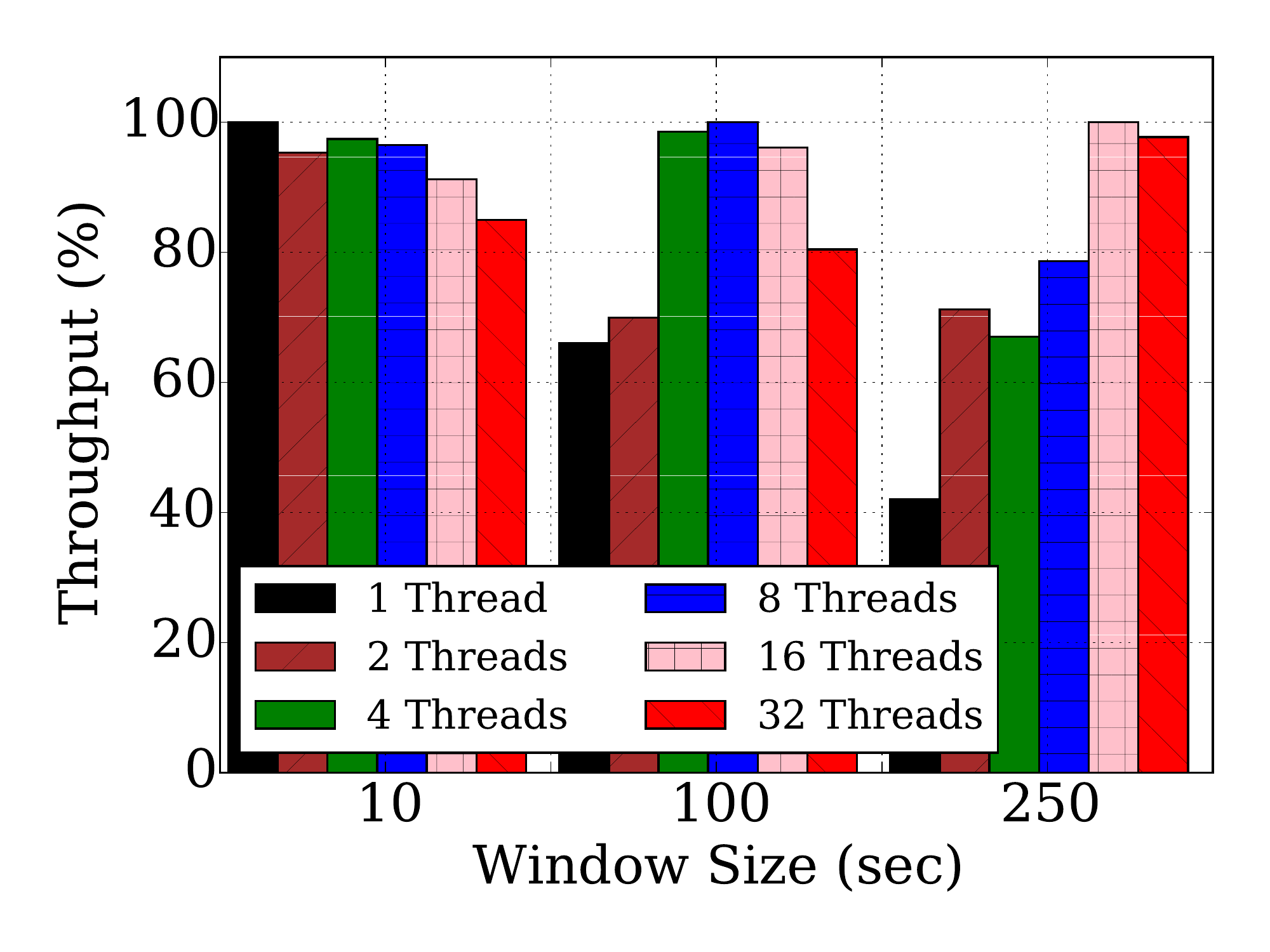}}
	\vspace{-8pt}
	\caption{Throughput evaluations for different window sizes on shared memory infrastructure.}
	\label{fig:evalsOctopus}
	\vspace{-10pt}
\end{figure*}

\section{Evaluations}
\label{sec:evaluation}

In this section, we present our experiments with StreamLearner on the DEBS Grand Challenge 2017 data set with 50,000 sensor data events. 

\textbf{Experimental Setup: }
We used the following two computing environments. (i) A notebook with $4 \times 3.5$ GHz (8 threads, Intel Core i7-4710MQ), $8$ GB RAM (L1 Cache 256 KB, L2 Cache 1024 KB, L3 Cache 6144 KB), and 64 Bit support. 
(ii) An in-house shared memory infrastructure with $32 \times 2.3$ GHz (Quad-Core AMD Opteron(tm) Processor 8356), and $280$ GB RAM (L1d cache 64 KB, L1i cache 64 KB, L2 cache 512 KB, L3 cache 2048 KB), and 64 Bit support.

\textbf{Adapting the window size $W$: }
In Figure~\ref{fig:scalabilityNotebook}, we show the absolute throughput of StreamLearner on the y-axis and different window sizes $W$ on the x-axis using the notebook for a different number of threads. Clearly, larger window size leads to lower throughput as computational overhead grows.
We normalized this data in Figure~\ref{fig:scalabilityNotebookBar} to the interval $[0,100]$ to compare the relative throughput improvements for the different number of threads. Clearly, the benefit of multi-threading arises only for larger window sizes due to the constant distribution overhead that can not be compensated by increased parallelism because each thread has only little computational tasks between points of synchronization (on the splitter and on the merger).
Overall scalability is measured in Figure~\ref{fig:scalabilityNotebook}. It can be seen that StreamLearner scales best for data-parallel problems with relatively little synchronization overhead in comparison to the computational task. For small window sizes (e.g. $W=10$), throughput does not increase with increasing number of workers. However, for moderate to large window sizes, scaling the number of worker threads has an increasing impact on the relative throughput: scaling from one to nine threads increases throughput by $2.5 \times$.

In Figure~\ref{fig:scalabilityOctopus}, we repeated the experiment on the shared-memory infrastructure. The first observation is that the single threaded experiments are four times slower compared to the notebook infrastructure due to the older hardware. Nevertheless, in Figure~\ref{fig:scalabilityOctopusNormalized}, we can see clearly that the relative throughput decreases when using a low rather than a high number of threads (e.g. for larger window sizes $W>100$). In Figure~\ref{fig:scalabilityOctopusBar}, we measure scalability improvements of up to $60\%$. Nevertheless, it can be also seen that it is not always optimal to use a high number of threads -- even if the problem is highly parallelizable.

\begin{figure}
	\includegraphics[width=0.7\linewidth, clip=true, trim=0pt 0pt 0pt 0pt]{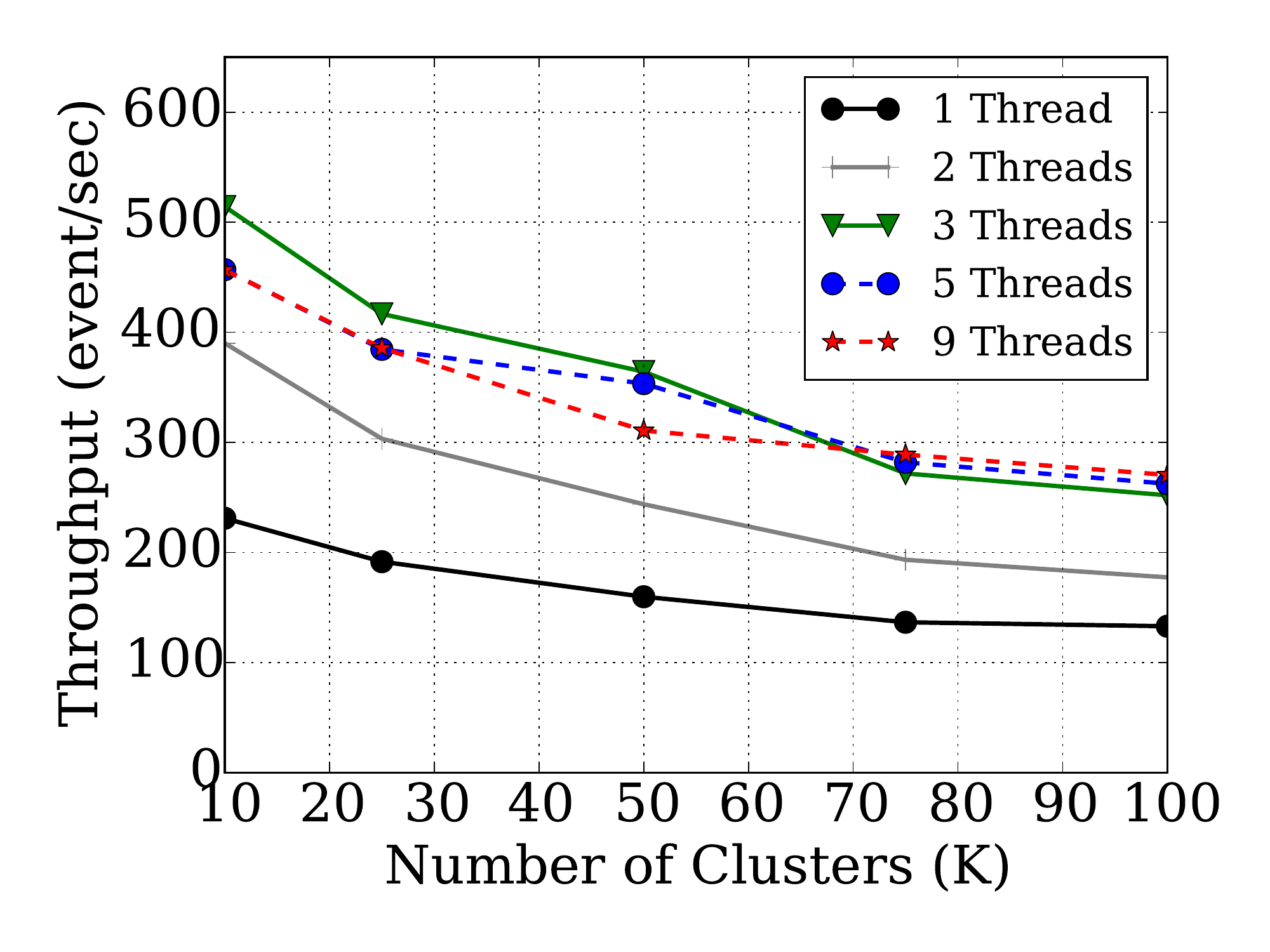}
	\vspace{-10pt}
	\caption{Throughput for varying number of clusters $K$.}
	\label{fig:evalsK}
	\vspace{-10pt}
\end{figure}

\textbf{Adapting the number of clusters $K$: }
In Figure~\ref{fig:evalsK}, we plot the absolute throughput for a varying number of clusters and different threads. We fixed the window size to $W=100$. Not surprisingly, an increasing number of clusters leads to reduced throughput due to the increased computational complexity of the clustering problem.
Evidently, increasing the number of threads increases the throughput up to a certain point. This is consistent with the findings above.

\section{Conclusion}
\label{sec:conclusion}

StreamLearner is a distributed CEP system and API tailored to scalable event detection using Machine Learning on streaming data.
Although our API is general-purpose, StreamLearner is especially well-suited to data-parallel problems -- with multiple event sources causing diverse patterns in the event streams. For these scenarios, StreamLearner can enrich standard CEP systems with powerful Machine Learning functionality while scaling exceptionally well due to the pipelined incremental training and inference steps on independent models.

\bibliographystyle{ACM-Reference-Format}
\bibliography{bib/bib} 

\end{document}